# Beating of grafted chains induced by active Brownian particles

Qiu-song Yang,　Qing-wei Fan,　Zhuang-lin Shen,　Yi-qi Xia,　Wen-de Tian[*],　Kang Chen[*]

Center for Soft Condensed Matter Physics & Interdisciplinary Research, College of Physics, Optoelectronics and Energy, Soochow University, Suzhou 215006, China

Corresponding authors: tianwende@suda.edu.cn (W. T); kangchen@suda.edu.cn (K.C.);

**Abstract**

　　We study the interplay between active Brownian particles (ABPs) and a "hairy" surface in two dimensional geometry. We find that the increase of propelling force leads to and enhances inhomogeneous accumulation of ABPs inside the brush region. Oscillation of chain bundles (beating like cilia) is found in company with the formation and disassembly of dynamic cluster of ABPs at large propelling forces. Meanwhile chains are stretched and pushed down due to the effective shear force by ABPs. The decrease of the average brush thickness with propelling force reflects the growth of the beating amplitude of chain bundles. Furthermore, the beating phenomenon is investigated in a simple single-chain system. We find that the chain swings regularly with a major oscillatory period, which increases with chain length and decreases with the increase of propelling force. We build a theory to describe the phenomenon and the predictions on the relationship between period and amplitude for various chain lengths and propelling forces agree very well with simulation data.

**Introduction**

Active matters are able to take energy from environment to drive their motion far from equilibrium.[1] In recent decades, they have been extensively studied as a subject of non-equilibrium statistical physics.[2] A group of self-propelling agents can exhibit rich and intriguing emergent phenomena, such as nontrivial fluctuations and pattern formation, from microscopic to macroscopic scale.[2,3] Varieties of artificial ABPs have been made in recent years and they become one of the major types of active matters.[4] Their non-equilibrium motion is featured by active swimming and random fluctuation due to solvent. A suspension of ABPs shows distinct collective behaviors in contrast to its passive analog, such as clustering or motility-induced phase separation (MIPS) and active micro-rheology.[5–8]

Recently, the behavior of ABPs encountering with passive objects or boundary has attracted many interests.[6,9–15] Hard surface can be used to efficiently manipulate the distribution of ABPs[10]. For example, the steady-state density distribution of particles along the boundary is proportional to the local curvature[10]; self-propelling rods can be completely captured by a wedge at certain apex angles.[16] In biological world, soft surface is ubiquitous.[17] ABPs can exert heterogeneous mechanical pressure on soft surface and disturb its shape.[18] A soft vesicle enclosed with ABPs was found to expand significantly. Its shape alters dynamically and becomes elongated when the propelling force is large and the enclosed particles are not so dense.[19,20] Another interesting example is a soft-chain-grafted colloidal disk immersed in the bath of ABPs. Positive-feedback-like cooperation between particle trapping and collective chain deformation leads to the longstanding spontaneous symmetry breaking and hence the unidirectional rotation of the disk.[21]

Brush-like surface is ubiquitous in both nature and industrial applications. For example, polymer brush is widely used to modify or improve the properties of a surface for desired chemical affinity, and biocompatibility.[22,23] Array of cilia is an example of brush-like structures in biological world. Our previous work has focused on the statistical behavior of brush deformation and active particle penetration in three dimensions (3D).[24] In this work, we study the interplay between brush-coated surface and ABPs in two dimensions (2D), which may happen in confined geometry, e.g. thin film or liquid-liquid interface. With the reduction of dimension, the steric interaction exaggerates the coupling between the motions of ABPs and grafted chains. We find that, in company with the formation and disassembly of dynamic clusters of ABPs, the chains form bundles and swing periodically in analogous to the beating of cilia. The larger the propelling force is, the larger the swinging amplitude of the bundles will be. Single-chain simulation shows that the swinging period (frequency) increases nearly linearly with chain length (propelling force).

**Model and Simulation Methods**

In our simulation, a grafted chain is modeled as a sequence of bond-connected beads with one of its ends attached to a flat substrate. The substrate is mimicked by smooth Weeks-Chandler-Andersen (WCA) wall,[25] which prevents the components in the system from passing through the substrate. The chains are uniformly grafted onto the substrate with a line density, $0.3/\sigma$. Each chain is composed of $N_p$ beads. An ABP is treated as a self-propelled disk driven by a force $F$ along the direction $\hat{u}(t)$, which reorients under fluctuation. Each particle or bead has a mass, $m$, and diameter, $\sigma$.

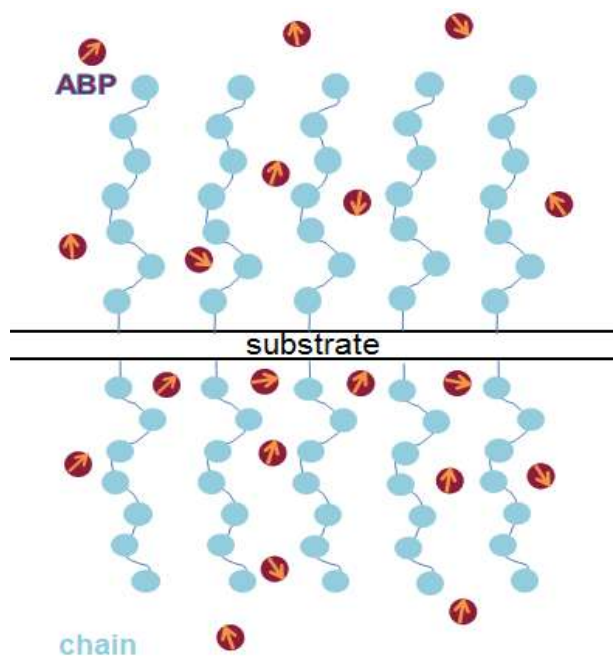

Fig.1. A schematic of the system. Bead-spring chains are grafted on the substrate. The ABPs are modeled as self-propelled disks with the propelling force along their individual inherent orientation.

The purely repulsive WCA potential is also adopted for the non-bonded interactions between all beads and active particles.

$$U_{WCA}(r) = \begin{cases} 4\varepsilon\left[\left(\frac{\sigma}{r}\right)^{12} - \left(\frac{\sigma}{r}\right)^{6}\right] + \varepsilon, & r < r_c = \sqrt[6]{2}\sigma \\ 0, & r > r_c \end{cases} \quad (1)$$

$\varepsilon$ is the interaction strength. The harmonic potential of successive beads is $U = k(r - r_0)^2$. $k$ is the spring constant, $r$ the distance between the bonded beads and $r_0$ the equilibrium bond length. We set $k = 1000\varepsilon/\sigma^2$ and $r_0 = 0.98\sigma$.

The Langevin equation is used to describe the motion of particles and beads,

$$m\ddot{\boldsymbol{r}}_i = -\frac{\partial U_i}{\partial r_i} - \gamma \dot{\boldsymbol{r}}_i + F\hat{u}_i(t) + \eta_i(t) \quad (2)$$

$$\dot{\theta}_i = \xi_i(t) \quad (3)$$

where $\boldsymbol{r}_i$ is the position of the $i$-th entity. $\eta_i(t)$ is a Gaussian white noise induced by implicit solvent, which satisfies the fluctuation-dissipation theorem, $\langle \eta_{j,\alpha}(t)\eta_{l,\beta}(t') \rangle = 2D_t\delta_{jl}\delta_{\alpha\beta}\delta(t - t')$,[26] where $\alpha$ and $\beta$ denote components of Cartesian coordinates, $D_t$ the translational diffusion constant. The translational friction coefficient $\gamma = \frac{k_BT}{D_t}$. Additionally, the rotational noise $\xi_i$ is also Gaussian with zero mean and $\langle \xi_j(t)\xi_l(t') \rangle = 2D_r\delta_{jl}\delta(t - t')$, where $D_r$ is the rotational diffusion constant.

Equation (2) governs the translational motion. For the chain beads, only Eq. (2) is required with $F = 0$, and $U_i$ is composed of both non-bonded WCA potentials and bonded harmonic potentials. But for active particles, $U_i$ contains only the non-bonded WCA potentials. Eq. (3) is necessary to depict the coupled rotational kinetics of the driving direction.

We use the home-modified LAMMPS software to perform the simulations.[27] To simplify the boundary condition, we built the system symmetrically by grafting the chains on both sides of the substrate as shown in Fig. 1. Square box of $180\sigma \times 250\sigma$ with periodic condition in both x and y directions is adopted. Reduced units are used in the simulations by setting $m = 1$, $\sigma = 1$ and $k_BT = 1$. The corresponding time unit, $\tau = \sqrt{m\sigma^2/k_BT}$. As a general model of active Brownian particles, we treat $D_r$ and $D_t$ as two independent parameters[28], and set $D_r = 6 \times 10^{-4}$, which is small enough to manifest the activity of particles. Besides, we use $\varepsilon = 10$ and set the friction coefficient $\gamma = 10$ which is large enough that the motion is essentially overdamped.[29] For each case, it was run by a minimum time of $1 \times 10^7 \tau$ with a time step, $\Delta t = 2 \times 10^{-3} \tau$. The dimensionless propelling force is in unit of $k_B T/\sigma$.

**Results and discussion**

Structure and dynamics are the two aspects we concerned. Specifically, we try to understand: 1) how is the distribution of ABPs; 2) how is the cooperative motion of the grafted chains and the trapped ABPs in the brush region; 3) how is the statistical conformation of chains in the presence of ABPs?

**Distribution of ABPs.** Figure 2 shows the typical snapshots with the propelling force $F$ equal to 0.0, 0.5, 1.0 and 10.0, respectively. As expected, with the increase of propelling force, more and more ABPs accumulate in the brush region, the same as our observation in the 3D system.[24] On the one hand, excluded-volume interactions between particles and chains tend to keep the particles outside the brush region. On the other hand, the grafted chains lower the mobility of trapped ABPs, and hence cause the accumulation of particles in the brush region (enhanced self-trapping mechanism).[30] This effect dominates even at small propelling force (Fig. 2). Figure 3(a) shows an anomalous negative particle density gradient from the inside to the outside of brush region at $F \geq 0.5$.

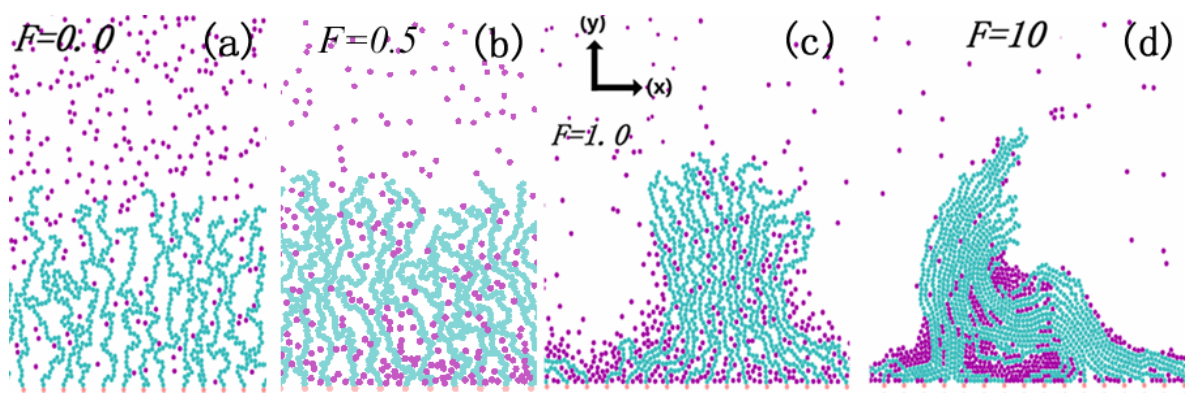

Fig. 2. Typical snapshots of grafted chains and ABPs for different propelling forces at chain

length $N_p = 60$.

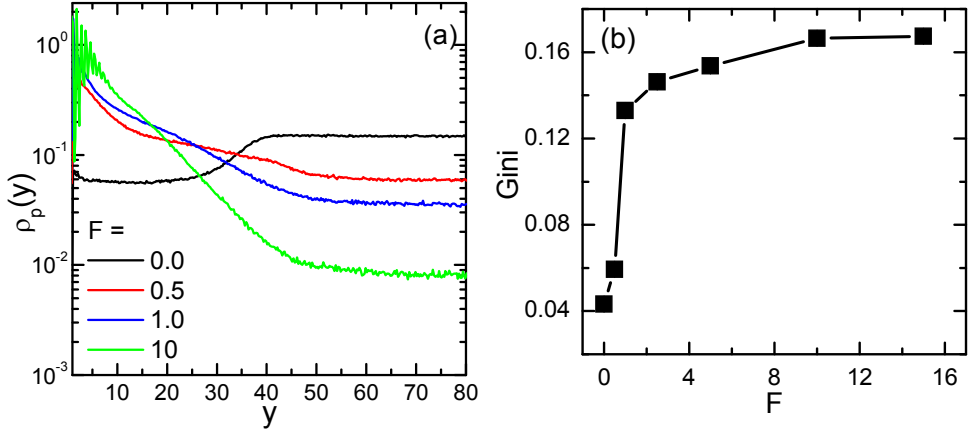

Fig. 3. (a) The density profile, $\rho_p(y)$, of ABPs for the upper half of the system. $y$ is the distance away from the smooth substrate. (b) Gini coefficient of ABPs along x direction as a function of propelling force (chain length $N_p = 60$).

An unexpected phenomenon, which is not observed in the corresponding 3D system[31] or the system of ABPs in contact with a flat wall, is the apparent inhomogeneous distribution of particles in the horizontal x direction (Fig. 2(c) and (d)). This inhomogeneous distribution of particles is accompanied by the formation of chain bundles. The formation of chain bundles creates free intervals in the brush region. However, the ABPs tend to pack around chains instead of staying in the intervals due to the enhanced self-trapping mechanism. As the propelling force increases, the chains in the bundle and the trapped particles pack more tightly. This emergent phenomenon is a consequence of the enhanced steric effect in 2D which brings the strong coupling between the motions of ABPs and passive chains. To quantify such inhomogeneous distribution of particles, we calculate the Gini coefficient along the x direction. We partition the simulation box into 50 slices along the x direction. The Gini coefficient[28] is defined as $Gini = \frac{1}{2\bar{\rho}}\sum_i \sum_j |\rho_i - \rho_j|$ where i and j are the slice indices, $\rho_i$ the number density of particles in the $i$-th slice, and $\bar{\rho}$ the mean number density. Larger Gini coefficient means stronger inhomogeneous distribution of particles. Figure 3(b) shows that the horizontal inhomogeneity of particle distribution initially grows rapidly with propelling force and then gradually approaches a plateau. In contrast, the Gini coefficient of the horizontal distribution of ABPs in the 3D brush system[31] is small and almost independent of the propelling force, suggesting homogeneous horizontal distribution of ABPs (Fig.S1 in the Supporting Information). The physics underlying such difference is the enhanced self-trapping effect in 2D geometry.

**Assembly of particles and beating of chain bundles**. An intriguing phenomenon observed in our 2D system is the formation of chain bundles and beating. They are dynamic and unstable structures, which are also found in the system of a chain-grafted colloidal disk immersed in the bath of ABPs.[21] The mechanism of the formation of these bundles is closely related to the phase separation in mixtures of active and passive particles[32] and/or the dynamic depletion attraction found between two parallel hard rods in a 2D active particle bath.[33] The bundle swings from one side to the other driven by the trapped ABPs on its two sides (see Fig. 4, the movie SI_1_movie and Fig. S2 in the supporting materials). In Fig. 4, the chain bundle is initially bent to the left. The left side of the bundle is concave and the trapped ABPs on this side are tightly packed as a cluster. While the trapped ABPs are loosely distributed on the right convex side of the bundle. Thus, the imbalance of the active pressure on the two sides of the bundles from the trapped ABPs drives the motion of the bundle to the right. During this process, the tightly packed cluster on the left disassembles, and particles gradually spread along the side of the bundle. Some of them on the top part of the bundle eventually leave the brush region. On the contrary, the right side of the bundle gradually changes from convex to concave. More ABPs become trapped in this high-curvature region and assemble into a cluster. Therefore the motion of the bundle finally reverses. This process repeats and hence the beating of chain bundles occurs. Figure 4 and the supporting movie SI_1_movie only show a selected short period of one beating bundle in a large system. Actually the swinging bundles in the 2D strongly interfere each other, especially for high grafting density and long chain (Fig.S2), which leads to the irregularity of the beating. The mechanism of this beating is different from that of the natural cilia or flagella, which originates from the tangential force by inside motor proteins. It is also different from the fluid-flow-induced beating reported by Sarka and Thakur[34], who found that two competing forces are necessary for the occurrence of the oscillation, i.e. the elastic force due to polymer rigidity and the active force due to chemical activity.

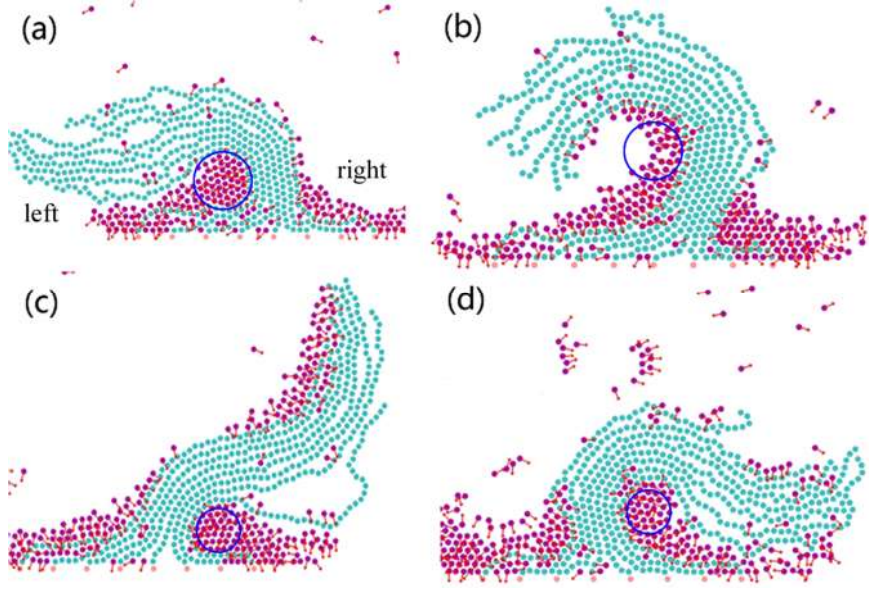

Fig. 4 Time evolution (from a to d) of the configurations of a beating chain bundle and trapped ABPs for $F = 10.0$, $N_P = 60$. The arrows denote the velocity directions of ABPs. Blue circles represent the concave regions.

**Chain statistics.** The statistical properties of the brush layer in the bath of ABPs are anticipated to be different with its analog in the passive bath. Here we first calculate the average contour length which reflects the mean stretch or tension of the chains. Figure 5(a) shows that the mean contour length increases roughly linearly with the propelling force. Because of the large elastic constant of the bonds, the extension of the chain length is slight in the range of propelling forces that we explore. But simple calculations indicate that the tension along the chain is numerically close to the value of the propelling force. The tension mainly comes from the drag on the chains by trapped ABPs which induces an equivalent shear stress.[18] Under the drag, the chains/bundles not only stretch but also being pushed down. This can be manifested by calculating the average height of the chains, which is defined as $\langle h \rangle = \frac{\int \rho(y) y dy}{\rho(y) dy}$. Figure 5(b) gives the height, $\langle h \rangle$, as a function of propelling force. It shows an apparent increase of $\langle h \rangle$ at $F = 0.5$, reflecting the vertical stretch of chains due to the enhanced exclude-volume repulsion (see Fig.2). And then the height decreases with the propelling force corresponding to the tilting of chains to form bundles and their beating at large $F$. From the trajectory (not shown), we find that the chains do not form bundles at $F = 0.5$. They form bundles with very weak beating phenomenon at $F = 1.0$ and form tight bundles and beat significantly at $F \geq 2.5$. In contrast, the excluded-volume repulsion in the 3D system always stretch the chains, so that the brush height mainly increases with propelling force and only slightly decreases at large propelling force.[24] The novel statistical behavior of the brush in 2D can also be manifested by the terminal distribution (Fig. 5(c)). The peak position of the free ends is around $y = 35$ in the case of passive particles, i.e. $F = 0$. It shifts to higher values for the propelling forces $F = 0.5$ and $1.0$, indicating the vertical stretch of chains. The decrease of peak value for $F = 1.0$ corresponds to the formation of chain bundles. For large $F \geq 2.5$, the distribution becomes very broad and the peak position shifts to smaller y, in response to the beating of chains.

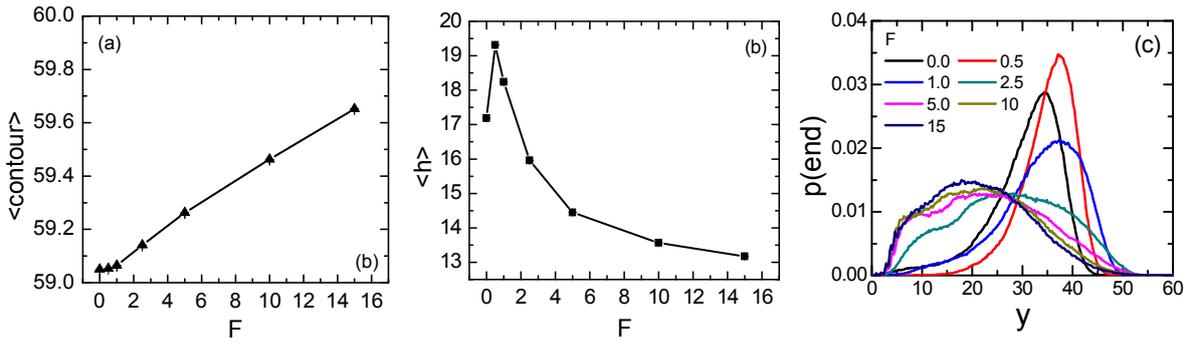

Fig. 5. The average contour length (a) and the average height $\langle h \rangle$ (b) as functions of propelling force. (c) The probability distribution of free terminals of chains. ($N_P = 60$)

**Beating of a single chain.** Here we investigate the beating phenomenon in the simplest case, i.e. there is only one chain grafted on the substrate, which is equivalent to a very dilute grafting situation. Surprisingly, the beating of the chain turns out to be very regular. To quantify the periodicity of beating, we measure the displacement of the free end of the chain relative to the grafted point in the x direction. Figure 6(a) (and the supporting movie SI_2_movie) shows the regular periodic beating of the single chain at F=10. The spectrum analysis in Fig. 6(b) proves the existence of a dominant frequency, $1.6 \times 10^{-3}/\tau$. The basic beating mechanism and behavior of a single chain are analogous to that of a bundle. The main differences are the stability and regularity. In the absence of steric hindrance by other chains, the amplitude of the single-chain beating is much larger and the main body of the chain can keep lying down when the propelling force is large. As a result, the curvature of the chain at the place in contact with the moving particle cluster almost does not change during beating and correspondingly the particle cluster keeps packed tightly until reaching the free end.

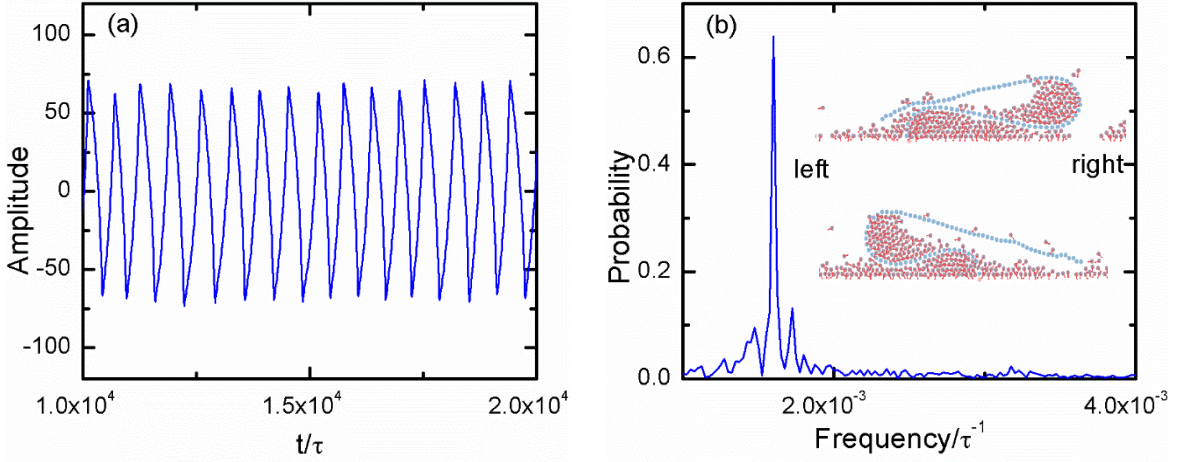

Fig. 6. (a) Time sequence of the $x$-displacement of the free end relative to the grafted point. (b) The corresponding spectrum analysis. ($F = 10$ and $N_P = 80$)

Chain length and the propelling force are two important factors that influence the beating phenomenon. Figure 7(a) shows the chain-length dependence of the main period $T$ and the amplitude $A$ of the beating for $F = 10$. Both $T$ and $A$ increase roughly linearly at large chain lengths. Since the chain almost lies horizontally except the part close to the grafted point (see the insets of Fig. 6(b)), $A$ should numerically close to (slightly smaller than) $N_p$. This is verified by the ratio $A/N_p$, which increases and approach 1 with the increase of $N_p$ (inset of Fig. 7(a)). We also vary the propelling force at fixed chain length, $N_p = 60$ (Fig.7 (b)). The amplitude $A$ exhibits a significant increase at small force and then approaches a plateau at large force. This is consistent with the picture that the main part of the chain almost lies horizontally at large force. $N_p - A$ approaches the plateau value around 10, corresponding to the part of the chain which is close to the grafted point and nearly orients vertically. The frequency of beating turns out to be a merely linear function of the propelling force when $F$ is large.

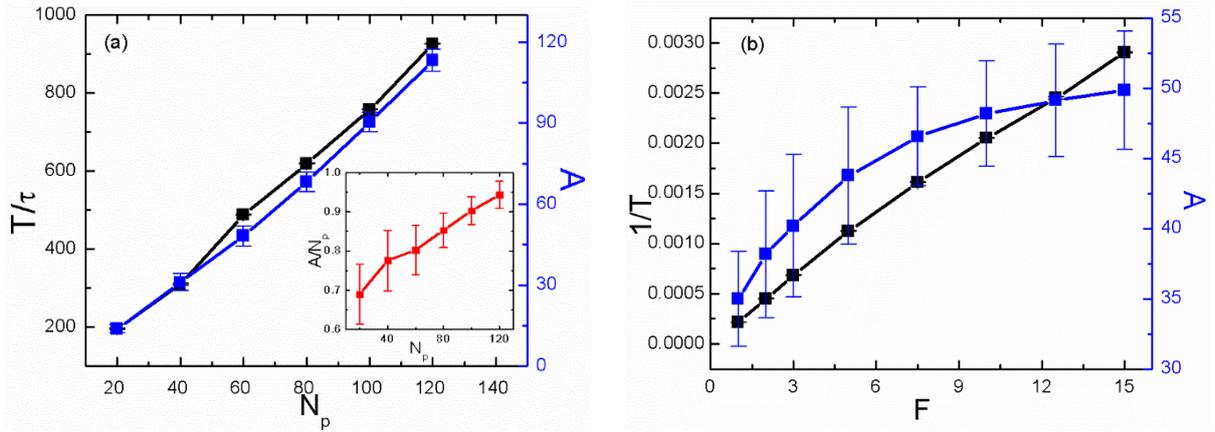

Fig. 7. (a) Variation of period $T$ and amplitude $A$ of the beating of a single grafted chain with chain length. The inset shows the variation of the ratio, $A/N_p$. ($F = 10$) (b) Variation of frequency $1/T$ and amplitude $A$ with propelling force. ($N_p = 60$)

To understand the physical process, we build a simple theory for the situation of large propelling force and modest or large chain length. In this situation, the main part of the chain lies almost horizontally during beating and the particle cluster that pushes the chain to move keeps packed tightly. In the beating, the trapped particle cluster and the part of chain above it are in motion. Phenomenologically, the motion is a reminiscent of a pulley system (see Fig. S3 in the supporting information). In the horizontal direction, the speed of the upper part of the chain is twice the speed of the particle cluster. We establish a horizontal coordinate with the origin at the grafted point (the inset of Fig. 8(a)). $x$ describes the position/displacement of the moving particle cluster and hence $dx/dt$ is its speed. The upper part of the chain then moves with a speed of $2dx/dt$. The number of beads of the upper part of the chain is approximately $(A - x)/r_0$ (The change of bond length is negligible (Fig. 5(a))). The number of ABPs in the moving cluster $n_c$ varies with $N_p$ and $F$ as shown in Fig. 8(a). It increases drastically with $N_p$, while quickly saturates at large F. In the overdamped limit, the "pulley system" is in balance, i.e. the effective driving force equals the total drag force on the chain/particle-cluster assembly. Therefore, in the horizontal ($x$) direction, we have the equation:

$$2 \cdot \gamma \frac{A-x}{r_0} \cdot 2 \frac{dx}{dt} + \gamma n_c \frac{dx}{dt} = \alpha n_c F \qquad (4)$$

The first term on the left side of Eq. (4) is the drag force on the chain. The factor 2 in front comes from the assumption that the tensions in the upper and lower part of the chain are equal (see Fig. S3 in the supporting information). The second term on the left side of Eq. (4) is the drag force on the particle cluster. The term on the right side of Eq. (4) is the horizontal effective driving force due to the propelling force on each particle in the cluster. A straightforward assumption is that the effective driving force is proportional to the number of particles in the cluster and the magnitude of propelling force. $\alpha$ is a prefactor accounting for the average horizontal component of the propelling forces. Ideally, $\alpha = 2/\pi \approx 0.64$.[21] Here we treat $\alpha$ as a fitting parameter, which accounts for deviations due to simplification and other uncertainties. We integrate Eq. (4) within half period, i.e. t is from 0 to $T/2$ and $x$ from 0 to $A$ and obtain the following relation between $T$ and $A$:

$$T = \frac{2\gamma}{\alpha F}\left(\frac{2}{n_c}\frac{A^2}{r_0} + A\right) \tag{5}$$

With the knowledge of cluster size $n_c$ (Fig. 8(a)), we collect all the data of $T$ and $A$ that vary with chain length and propelling force from Fig. 7 and compare them with the theoretical predictions by Eq. (5) in Fig. 8(b). By setting the value $\alpha = 0.55$, the theory agrees very well with the simulation data, especially in the situation of large propelling force and long chain.

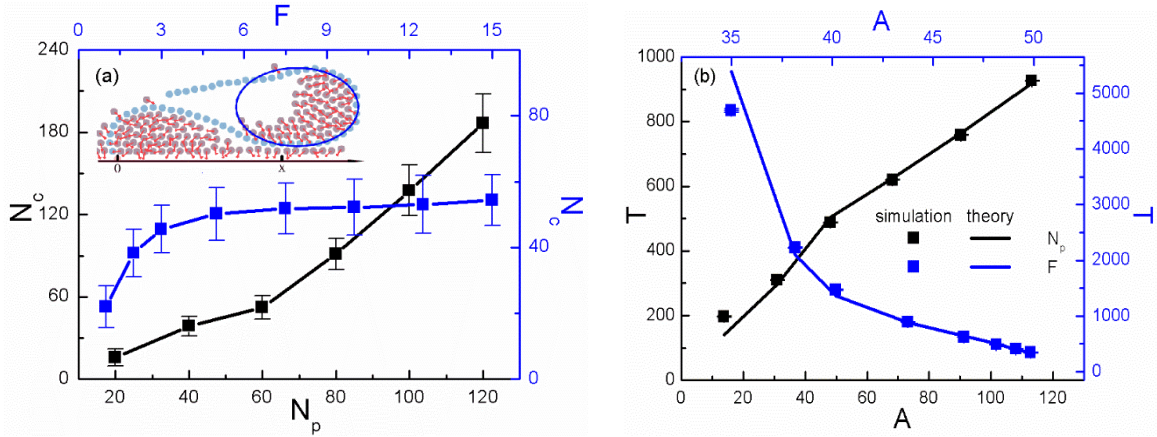

Fig. 8 (a) The moving cluster size $n_c$ as a function of $N_p$ and $F$. The inset shows that $n_c$ is the number of ABPs in the concave region (blue circle) of the chain. (b) The relationship between the period $T$ and amplitude $A$ of the beating for various $N_p$'s and F's. Theoretical predictions (solid lines) fitting with a single value, $\alpha = 0.55$, agree very well with the simulation data (squares).

**Summary**

We explore the beating phenomenon of grafted chains in contact with ABPs in two dimensions. The cause of this phenomenon is the coupling between the motions of ABPs and grafted chains, which is greatly exaggerated in 2D. Small propelling force facilitates the homogenous penetration of particles into the brush and hence increases the thickness of the brush layer due to the excluded-volume interaction. In contrast, the brush layer becomes unstable for large propelling force. The grafted chains form bundles and swings irregularly coupled with the formation, disassembly and motion of particle clusters. Statistically, the thickness of the brush layer decreases with propelling force even though more and more particles penetrate into the brush region. Meanwhile, the particle distribution becomes highly inhomogeneous. In the absence of interference from other chains, the single-chain beating turns out to be very regular. The oscillatory period and amplitude increase linearly with chain length. The inverse of period (frequency) increases linearly with propelling force, while the amplitude increases at small propelling force and saturates when the force is large. Finally, we build a simple theory which describes the single-chain beating quite well.

The synchronization of beating chains has been found in other systems such as semi-flexible active filaments[34] and tails of swimming sperms[35]. It was suggested that the hydrodynamic interaction plays an important role in the synchronous beating pattern. It is very interesting to check the possibility of beating synchronization in our system, when hydrodynamic interaction is included. Shape anisotropic biological active matter and artificial ABPs, such as active rods[36] and eccentric self-propelled colloids[37,38] are ubiquitous. One can imagine that active rods easily get stuck in the brush region due to its shape anisotropy and exert persistent force on the brush. This satisfies the basic requirement for the beating phenomena. The situation is more complex in the case of eccentric active particles. No straightforward speculation can be made. The kinetic behavior of the assembly of brush and eccentric self-propelled particles is an interesting topic for future.

**Supplementary Material**

See supplementary material for Fig. S1, Fig.S2, and Fig.S3. Additionally, two supporting movies were provided for bundle beating (SI_1_movie) and single chain beating (SI_2_movie) at $F = 10.0$, $N_P = 60$.

**Acknowledgment**

This work was supported by the National Science Foundation of China (NSFC). W. Tian acknowledge financial support from NSFC Nos. 21474074 and 21674078. K. Chen acknowledge financial support from NSFC Nos. 21574096, 21774091, and 21374073.